\documentstyle[psfig,12pt,aasms4]{article}
\def\zabs{$z_{\rm abs}$}
\def\zem{$z_{\rm em}$~}
\def\lya{Ly$\alpha$ }

\def\nv{N~{\sc v}~ }

\def\cii{C~{\sc ii}~}

\def\civ{C~{\sc iv}~}
\def\civa{C~{\sc iv}$\lambda$1548~ }
\def\civb{C~{\sc iv}$\lambda$1550~ }

\def\siiv{Si~{\sc iv}~}

\def\kms{$km~s^{-1}$}

\begin{document}
\title{\bf High resolution study of associated \civ absorption systems in NGC
5548\altaffilmark{1}
}
\author{R. Srianand}
\affil{IUCAA, Post Bag 4, Ganesh Khind, Pune 411 007, 
India \\
anand@iucaa.ernet.in\\}
\altaffiltext{1}{Based on observations made with the NASA/ESA Hubble Space Telescope, obtained from the data archive at the Space Telescope Science Institute. STScI is operated by the Association of
Universities for Research in Astronomy, Inc. under NASA contract NAS 5-26555.}

\begin{abstract}
We present the results of a careful analysis of associated absorption
systems toward  NGC 5548. 
%We show most of the well resolved narrow
%components in the associated systems are line-locked with velocity
%separation equal to the \civ doublet splitting. 
Most of the well resolved narrow components in the associated system,
defined by the \lya, \civ and \nv profiles, show velocity separation
similar (to within 10~\kms) to the \civ doublet splitting. We estimate
the chance probability of occurrence of such pairs with velocity separation
equal to \civ doublet splitting
to be $6\times10^{-3}$. Thus it is
more likely that most of the narrow components are line-locked with
\civ doublet splitting.  This will mean that the radiative acceleration
plays an important role in the kinematics of the absorbing clouds. We
build grids of photoionization models and estimate the radiative
acceleration  due to all possible bound-bound transitions.  We show
that the clouds producing absorption have densities less than
$10^9~cm^{-3}$, and are in the outer regions of the broad emission line
region (BLR). We note that the clouds which are line-locked can not
produce appreciable optical depths of O~{\sc vii} and O~{\sc viii}, and
hence can not be responsible for the observed ionized edges, in the
soft X-ray.  We discuss the implications of the presence of optically
thin clouds in the outer regions of the BLR to the models of broad
emission lines.

\end{abstract}

\keywords{galaxies: active -- galaxies: individual (NGC 5548) --
galaxies: seyfert -- quasars: absorption lines}

\section{Introduction}

Associated absorption systems (i.e. absorption systems with
\zabs$\simeq$\zem), seen in $\sim10\%$ of the QSOs (and $\sim 50\%$ of
low $z$ AGNs), are a very good probe of the gas in the inner regions of
the QSOs (AGNs). Studies of associated systems provide clue toward (i)
understanding the chemical enrichment history (Petitjean et al. 1994;
Hamann, 1997; Petitjean \& Srianand, 1999), (ii) investigating the gas
dynamics (infall/outflow), and (iii) probing the geometry and velocity
field in the broad emission line region (BLR) (Srianand \&
Shankaranarayanan, 1999).  There are other ways of probing the gas in
the centres of AGNs/QSOs.  Broad emission lines and their
correlated variability with the ionizing continuum probe the emission
line gas (Peterson, 1993). Various observations suggest that in most of
the low redshift, low luminosity AGNs the BLR consists of optically
thick as well as optically thin clouds contributing to the total
emission (Shields et al. 1995). It is realized that the ionization
conditions in the optically thin clouds resembles that of clouds
producing associated absorption lines (Shields et al.  1995).
Ionization edges due to photoionized gas in the soft X-ray range (due
to O~{\sc vii}, O~{\sc viii} etc.,) are detected in large fraction of
AGNs (Reynolds, 1997). These systems are known as "warm absorbers".
The studies of "warm absorbers" probe a very highly ionized gas
components along the line of sight.

Modeling of different components discussed above requires a good
knowledge of the shape of the ionizing radiation, location, chemical
composition and density of the cloud.  Within the allowed range of
different parameters it is realized that one can get a unified model
for the absorbing cloud which will produce the observed optical depths
of ionized edges in soft X-ray and UV absorption lines (Mathur et al.,
1995). High resolution observations of some of the nearby AGNs show
that the associated absorption systems are made up of number of narrow
components (Crenshaw et al. 1998). However the X-ray spectra used to
detect the "warm absorbers" are of very low resolution and it is
impossible to establish any firm connection between the individual
components in the UV absorption and the ionized edge seen in the
X-rays. However, the important things one need to understand are (1)
location of the UV absorbers, (2) kinematics of the clouds, (3) how much
contribution such clouds will make to the emission line? and (4) how
does the clouds respond to the continuum variation?.

In this work we concentrate on the associated absorption systems
observed along the line of sight to NGC 5548.  We investigate the
multi-component velocity structure in the associated system using the
HST archival data. The absorption line profiles suggest that most of
the narrow components in the associated systems, identified using the
narrow unsaturated lines in \lya and \civ, show signatures of
line-locking with velocity separation equal to that of \civ doublet
splitting. We construct grids of photo-ionization models and estimate
the radiative acceleration due to line absorption in various
transitions.  We discuss the nature of the absorbing clouds using the
upper limit on the acceleration measured from the two epochs of
observations, constraints on the ionization parameter, $U$, (which is
the dimensionless ratio of number  density of ionizing photons to that
of particle density) and the requirements for the line-locking.

We show that the number density of the clouds is less than
$10^9~cm^{-3}$. This with the allowed range in $U$ suggests that the
absorbing cloud will be situated in the outer regions of the BLR or in
between BLR and NLR (Narrow emission line region). We also show that,
under the framework of models considered in our analysis, these
clouds can not produce appreciable optical depths of O~{\sc
vii} and O~{\sc viii}. Thus such clouds can not be responsible for the
ionized edges seen in NGC 5548.  We discuss the implications of
radiatively accelerated, optically thin clouds in the outer regions of
the BLR in the reverberation mapping studies.

\section{Sub-components in the associated system}

The combined FOS (Faint Object Spectrograph) spectrum of NGC 5548,
obtained for the AGN watch program (Korista et al. 1996), shows two
distinct components in the \lya absorption at 1232\AA (i.e \zabs =
0.0134) and 1234\AA (i.e \zabs = 0.0151)(Mathur et al., 1995).  The
velocity separation between these components  is $\sim500$ \kms, which
is very close to the \civ doublet splitting. These two components are
clearly present in the \civ profile in the GHRS (Goddard High
Resolution Spectrograph) spectrum (Mathur et al., 1999).  However the
low redshift system is weak and the profile is not well   defined due
to poor signal to noise. The very high spectral resolution ($\sim
46000$) achieved by the STIS (Space Telescope Imaging Spectrograph) data
allow us to perform a detail analysis of the individual velocity
components.  The GHRS and STIS spectra used in this study are retrieved
from the HST archive and the pipeline calibrated data are processed
using standard STSDAS tasks in IRAF.  The original data are published
already in the literature (Mathur etal ,1999; Crenshaw \& Kraemer, 1999).
     
\begin{figure}
\centerline{\vbox{
\psfig{figure=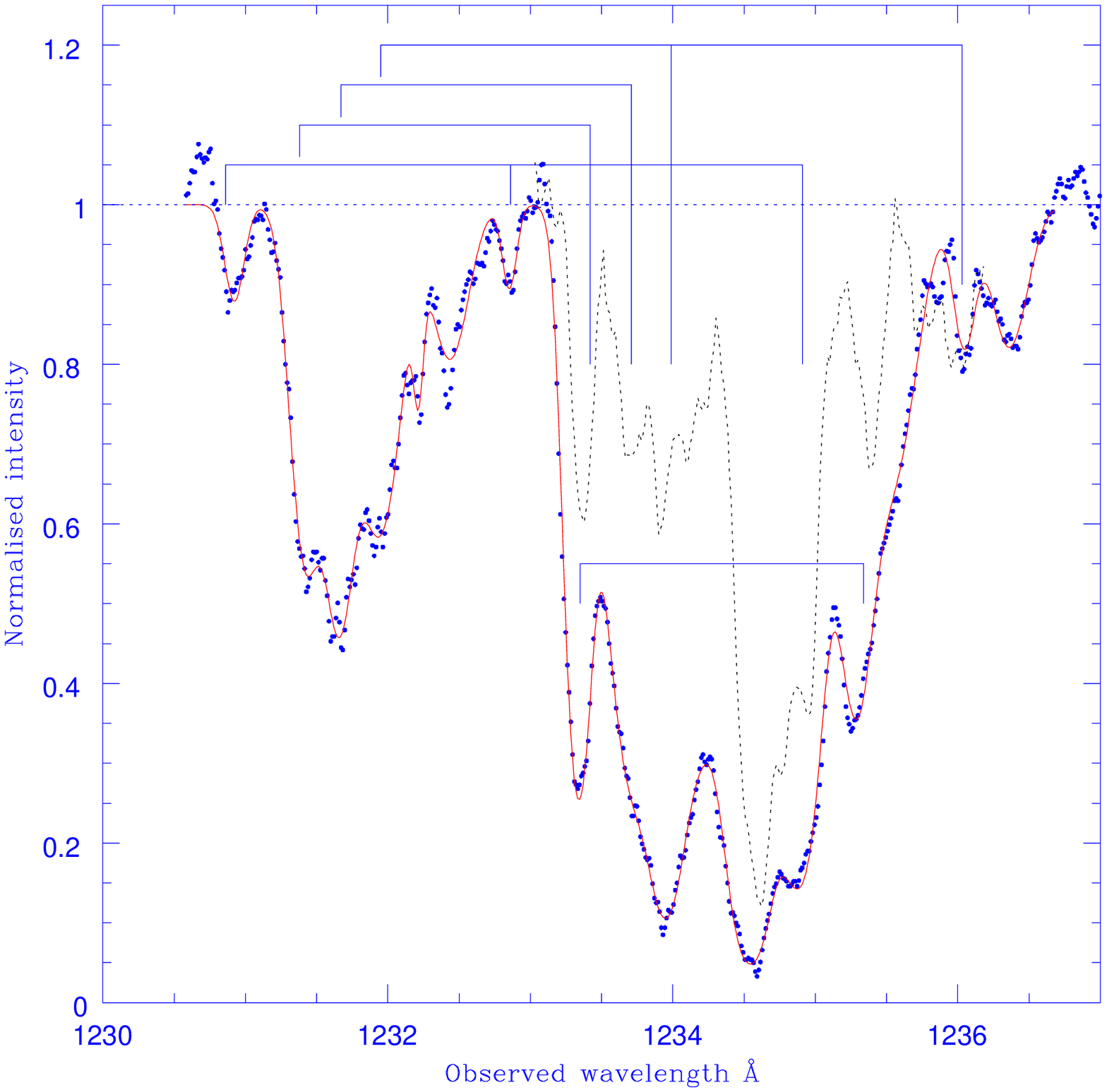,height=15.cm,width=15.cm,angle=0}
}}
\caption[]{\lya absorption line of the associated system toward NGC
5548. The dots and continuous curves represent the observed spectrum
and Voigt profile fits respectively. The dotted curve gives the profile
of the \civa line (suitably shifted to the \lya wavelength). Some of the
components which are separated by the \civ doublet splitting are also
marked in the
figure.}
\label{fig1}
\end{figure}

We plot the continuum normalised STIS spectrum (dots), in the
wavelength range of \lya absorption, in Fig~\ref{fig1}. As can be seen
from the figure that the sub-components in the two systems noted above are
clearly resolved.  The components in the low-$z$ system  are weak and
well resolved, while  the components in the high-$z$ system are strong
and blended. In order to resolve the components in the high-$z$ system
we make use of the \civ and \nv profiles from the GHRS and STIS data.
To illustrate the sub-components in this system we plot the \civ
profile (GHRS), shifted to the \lya wavelength range, using the dashed
curve.  We fit the absorption lines using a Voigt profile fitting code
(Khare et al., 1997). Minimum number of components needed to fit the blend is
decided using the $\chi^2$ per degrees of freedom.  The low-$z$ system
is resolved into 7 sub-components. As the continuum fit to the spectrum
is not unique (due to unknown intrinsic emission line profile) and  the
cloud producing the absorption need not cover the background source
completely, the derived column densities from our fits are lower
limits.  However the measurement of observed wavelengths of the
individual components are more reliable.  Individual components in the
low $z$ system have velocity dispersion in the range 15-40 \kms and a
typical error in the determination of the central wavelength is $\sim$
10 \kms.  The high-$z$ system is resolved using 11 components. The
centroid of each component is fixed to the value needed to fit the \civ
and \nv doublets from this system.  The velocity dispersion of the
individual components vary between 25 to 70 \kms and the error in the
wavelength is of the order of 7-10 \kms.

It is interesting to note that most of the components which are
resolved in the low-$z$ system have a corresponding components in the
high-$z$ system (as shown in Fig~\ref{fig1}) with a velocity splitting
corresponding to the \civ doublet splitting within an error of
10 \kms.  Also there are three components in the high-$z$ system which
have corresponding components with a velocity splitting corresponding
to the \civ doublet splitting among most of the resolved components.
Thus the \lya profile seen in the STIS
spectrum suggests the presence of line-locking with \civ velocity
splitting. We estimate the probability of occurrence of 8 pairs with
velocity splitting equal to that of \civ doublet splitting (within 10
\kms) among 18 clouds randomly distributed distributed in the velocity
range 0-1500 \kms~ to be  $6\times10^{-3}$. This suggests that the 
line locking seen is most probably true. Thus assuming the existence
of line-locking, in what follows we discuss various possible 
implications.

\begin{figure}
\centerline{\vbox{
\psfig{figure=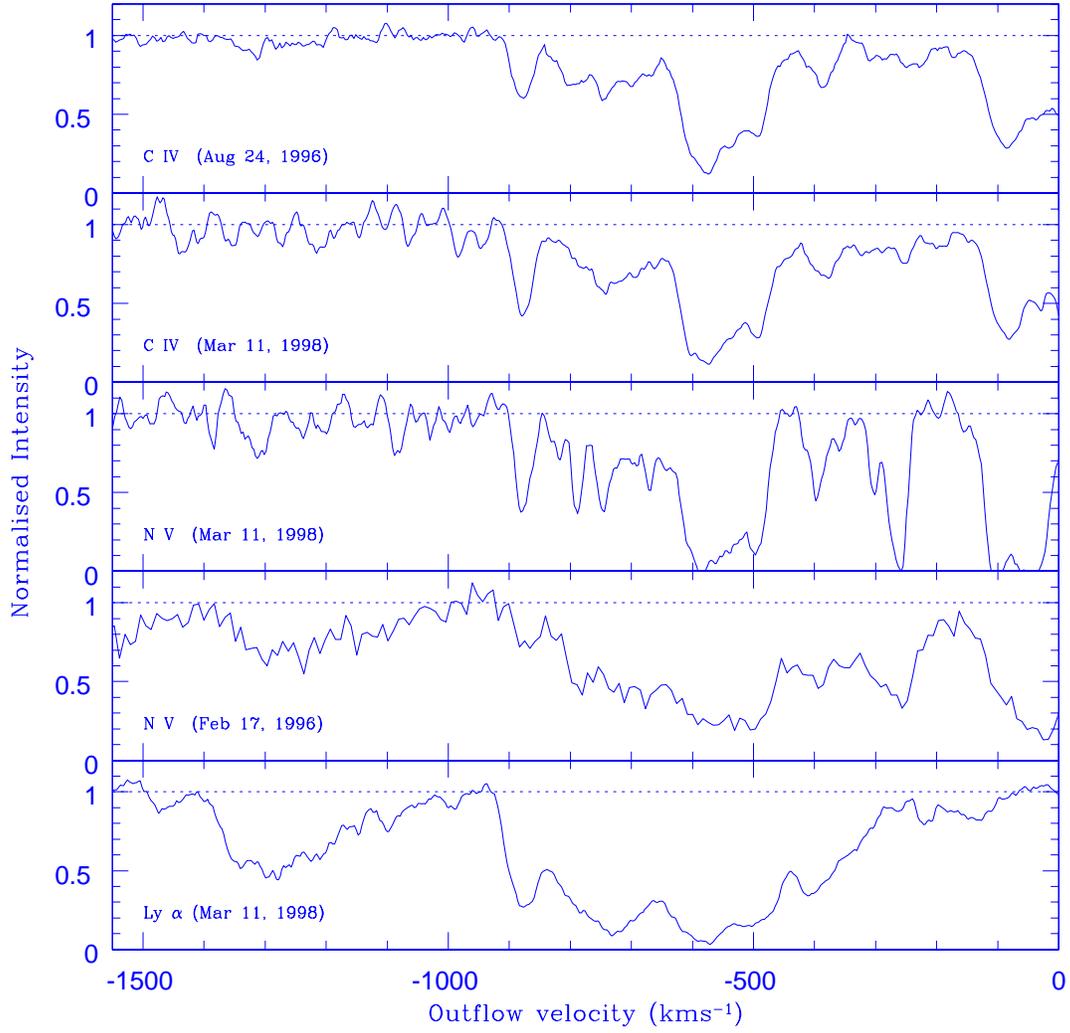,height=15.cm,width=15.cm,angle=0}
}}
\caption[]{ Profiles of \lya, \civ and \nv absorption lines
observed at different epoch. The outflow velocity is 
estimated assuming the redshift of NGC 5548 to be 0.0175.}
\label{fig2}
\end{figure}

The epoch of observations of NGC 5548 with GHRS and STIS are roughly 2
years apart. These observations can be used to get the bound on the
acceleration of the individual components. The velocity plot of \lya,
\civ and \nv absorption lines observed at different epoch are given in
Fig~\ref{fig2}. Apart from a uniform shift of -0.05 \AA~ applied to all
our GHRS data, we do not detect any shift in the wavelength of the
centroid of lines between the data observed in August 24, 1996 and
March 11, 1998. This gives a upper limit on the change in the velocity
of the components to be $\sim2\times10^{-2}~cm~s^{-2}$. Note that this
limit is strictly true for the components in the high-$z$ system
only.  In the case of low-$z$ systems poor signal-to-noise in the
expected region of \civ absorption in the STIS spectrum prevents us
from making any comparison between the \civ profiles.  Similarly the
GHRS spectrum in the \nv absorption region is noisy.  However, the \nv
profile resembles very well with the \lya profile observed with the
STIS. In the following sections we try to get the relationship for
radiative acceleration and acceleration due to gravity of a optically
thin (at the Lyman limit) photo-ionized cloud.

\section{Radiative Acceleration}

Transfer of momentum of photons from the central source to the
absorbing clouds through electron scattering, continuum ionization and
resonance line absorption can be used to accelerate the clouds.  If the
clouds are highly ionized, as in the case of associated \civ absorption
systems, the effect of continuum absorption is negligible.  Electron
scattering contributes to the acceleration only under high electron
column densities, i.e., $N_e\ge 10^{24}$ cm$^2$.  Thus in the range of
ionization parameters and electron density one needs to explain the
observed column densities of the associated systems the effect of
radiative acceleration due to electron scattering and continuum
absorption are negligible. Thus in this work we concentrate on the
radiative acceleration due to line absorption only.

It is the usual procedure to represent the ionization state of an
absorbing cloud using a dimensionless ionization parameter, $U$.  The
distance of the cloud from the continuum source, $r$, can be written
in terms of $U$ as,
\begin{equation}
r~=~\bigg({Q\over 4\pi U n_H C}\bigg)^{0.5}
\end{equation}
where $n_H$ is the number density of hydrogen (H~{\sc i} + H~{\sc ii}) atoms,  $C$ the velocity of 
light. $Q$, the number of ionizing photons emitted from the 
source is given by
\begin{equation}
Q~=~\int {L_\nu \over h \nu}~d\nu
\end{equation}
where $L_\nu$ is the luminosity at the frequent $\nu$. For a
given spectral energy distribution $Q$ can be estimated 
from the observed flux and a set of cosmological parameters.

The gravitational acceleration of a cloud towards the
central source is given by,
\begin{equation}
g~=~-GM~\bigg({4\pi U n_H C\over Q}\bigg)
\end{equation}
where $M$ is the mass of the central engine.

The total energy received by the cloud at any wavelength $\lambda$  per
unit wavelength interval per unit time is,
\begin{equation}
f~=~\bigg({L_\lambda A \over 4\pi r^2}\bigg)
\end{equation}
where $A$ is the area of the absorbing cloud.
Total momentum available in the radiation per sec per unit wavelength
interval at $\lambda$ is 
\begin{equation}
p(total)~=~\bigg({f\over C}\bigg).
\end{equation}

This momentum is imported to the absorbing cloud through resonance
line absorption and used to accelerate the cloud. The acceleration
due to a single line transition is given by,
\begin{eqnarray}
g_r^l&=&\bigg({L_\lambda A \over 4\pi r^2 mC}\bigg)\int_{line}1 -e^{\tau_\lambda} d\lambda\nonumber\\
     &=&\bigg({L_\lambda A U n_H \over mQ}\bigg)W^l
\end{eqnarray}
where, $\tau$, is the optical depth of the line $m$ is the mass of the
cloud and $W^l$ is the equivalent width to the line. If we assume
the absorbing cloud to be a plane parallel slab then,
\begin{equation}
g_r^l~=~\bigg({L_\lambda U n_H \over Q N_H m_H X}\bigg)W^l
\end{equation}
where, $n_H$, is the total hydrogen column density, $m_H$ is the mass
of hydrogen atom and $X$ is the mass fraction in terms of mass of
hydrogen atom. In order to get the total acceleration we need to sum 
the $g^l$  values due to all possible strong transitions. We assume
$X=1.33$ in all our models, and neglect the drag forces due to
ambient medium.

\section{Line-locking}

Line-locking is a process in which the velocity separation between
two absorption minima are equal to the velocity separation between some
allowed resonance transitions. If the flow is dominated by radiative
acceleration due to line absorption then the line-locking is achieved
through a non-local coupling process in which photons resonant with one
transition (at frequency $\nu_1$) in one part of the flow (say at a
location $x_1$, and velocity $v_1$) Doppler shift into resonance with
another transition (at frequency $\nu_2$) in another part of the flow
(say at $x_2$, and velocity $v_2$) with
\begin{equation}
v_2~=~v_1+{\nu_1-\nu_2 \over \nu_1}~C.
\end{equation}

In such a case the total number of photons at $\nu_2$ that can be
absorbed by the out-flowing material at $x_2$  is reduced due to
absorption at $\nu_1$ by the material at $x_1$. This reduces the net
acceleration at $x_2$. Thus the material at $x_2$ will spend some time
at an out-flow velocity corresponding to the velocity separation
between $\nu_1$ and $\nu_2$. In order to keep this stable for long time
two possible mechanisms are proposed.  
(1) {\bf Steady state locking:} In the case of steady state locking the
net acceleration drops to zero at $x_2$. This requires that either the
contribution by the line transition at $\nu_2$ dominates total
radiative acceleration - and this is normally not the case - or that the
total outward force is almost wholly counter balanced by an inward
force (eg. gravity or pressure drag) such that their difference is
smaller than the line radiation force due to the transition at $\nu_2$.
In order to achieve such a case one needs to fine tune the model
parameters. In this case the line-locking will occur at a definite
ejection velocities in the rest frame of the ionizing source (i.e. line
locking occurs on the absolute velocities).  The double absorption
troughs seen in \civ absorption in the case of few BALQSOs (Broad
Absorption Line Quasi Stellar Objects) suggest such an accelerating
mechanism is indeed important in some quasars (Weymann et al., 1991).
(2){\bf Non-steady state locking:} In this case, the locking is assumed
to be on velocity differences and not on absolute velocities relative
to the central source.  Locking occurs between two gas elements none
of which is at constant velocity (vanishing acceleration), it is only
the relative velocity between the two that remains constant and equal
to the doublet velocity. The fact that the line-locked pairs occur at
different velocities in the case of associated narrow absorption line
systems suggests that the non-steady state line locking is operating in
these systems. In this case no critical balance between the outward and
inward acceleration is required for locking to occur. What one needs is
the reduction is acceleration at $x_2$ due to lack of photons at
$\nu_2$ should be equal to the difference in the acceleration between
the two fluid elements before locking. While neither of the velocities
of the locking and locked gas, with respect to the central source, is
constant at locking their velocity difference is. Also locking can
occur in the in-falling as well as out-flowing clouds (Braun \& Milgrom,
1989).

\section{Parameters of the associated system in NGC 5548:}

In order to estimate the acceleration of the cloud one needs
to get the estimate of various parameters related to the AGN and
the absorbing cloud. In this section we try to get the allowed
bounds on different parameters from different available observations.
 
\subsection {Mass of the central engine:}

The mass of the central engine can be estimated assuming the width of
the broad emission lines are due to the gravitational influence of the
central engine. The distance of the BLR from the central engine is
obtained from the reverberation mapping studies.  Krolik et al. (1991),
based of the correlation between the line of sight velocity dispersion
of different emission lines and characteristic response time for each
line, estimated the mass of the central engine to be in the range
$10^7-10^8~M_\odot$.  Clavel et al. (1992) estimated the mass to be
3.7$\times10^7~M_\odot$, using the widths of the \lya line.
Done \& Krolik (1996), using the \civ emission line profile variability,
estimated the central mass to be 2$\times10^7M_\odot$ (for two
dimensional random motions) and 8$\times10^7M_\odot$ (for three
dimensional random motions) for the mass of
the central engine in the case of NGC 5548. Other method of estimating
the mass is using the accretion disk models and the continuum spectral
energy distribution. Sun \& Malkan (1989) need a minimum mass of
$2\times10^8~M_\odot$ to reproduce the energy distribution
 of NGC 5548 in the framework of standard disk models. However when
re-processing is added, the mass required reduces to
5.5$\times10^7~M_\odot$.  Thus we consider the mass to be
$10^7-10^8~M_\odot$ in our models for NGC 5548.

\subsection {Density in the absorbing clouds:} 

Shull \& Sachs(1993) have studied the associated \civ absorption in NGC
5548 using the IUE monitoring data. A clear anti-correlation between
the continuum luminosity and the \civ absorption line equivalent width
is visible in their data. Using the 4 days (sampling interval) limit on
the absorption line variability they estimated the electron density,
$n_e\ge5\times10^5~cm^{-3}$.  Mathur et al. (1995) have studied the HST
FOS spectra of NGC 5548 observed as a part of the AGN watch program
during 1993 with a sampling interval of $\sim$1 day.  They did not find
any change in the equivalent width of \civ absorption, within
measurement errors, in spite of a factor 1.5 variation in the UV
continuum flux. However they have noted that the data is consistent
with expected variability in the \civ equivalent width, within
measurement uncertainties. The high resolution studies of NGC 5548,
with GHRS and STIS, have revealed multi-component structure of the
absorbers. Also the absorption lines do not cover the background source
(BLR+continuum) completely and each component has different covering
factor.  The variability in the equivalent width might have some
contribution from the varying covering factor (Srianand \&
Shankaranarayanan, 1999). 

\subsection {Spectral energy distribution on NGC 5548:}

It is known that the optical-UV continuum in NGC 5548 is variable and
flattens when the object brightens.  However  there is little
simultaneous observations of NGC 5548 in various wave bands and the
change in the spectral energy distribution as a function of source
luminosity is unknown. Dumont et al.(1998) have provided detailed
discussion on spectral energy distribution of NGC 5548. In this study
we use the continuum used by Dumont et al (1998).  Assuming $q_o=0.5$,
$H_o = 75~kms^{-1} Mpc^{-1}$ and redshift of NGC 5548 to be 0.0175 we
estimate the value of $Q=3\times10^{54}s^{-1}$ from the flux in the
range 1 to 20 Ryd. Note that the spectrum we reconstruct from the
observations need not be the one seen by the absorbing clouds.  As
noted by Dumont et al. (1998), we should keep in mind that whole
continuum  distribution in the observed range is determined within
about a factor two of uncertainty ( and a large factor in the Extreme
UV range), owing to temporal variation.

\subsection {Ion column densities:}

Ultraviolet spectra of NGC 5548 observed with the Hopkins Ultraviolet
Telescope (Kriss et al. 1997) show no Lyman limit at the systemic
redshift. This suggest that $N(H{~\sc i})\le 10^{17}~cm^{-2}$ in
individual components.  The Voigt profile fits gives lower limit on the
$N(H{~\sc i})$ as we do not know the exact value of the covering
factor. The $N(H~{\sc i})$ in the individual components obtained from
the profile fits vary between $10^{13}$ to $2\times10^{14}~cm^{-2}$.
From the absence of \cii absorption we estimate the 2$\sigma$ upper
limit on $N(\cii)$ to be $10^{13}cm^{-2}$.  The lower limit on the \civ
column density derived from different components in the high-$z$ system
varies between 2$\times10^{13}$ to  $10^{14}~cm^{-2}$. Similarly, the
lower limit on the column density of \nv is in the range
2$\times10^{13}$ to 4$\times10^{14}~cm^{-2}$.  It is interesting to
note that \nv column densities in all the components are higher than
\civ. If the heavy element abundance ratios are like solar this will
mean high degree of ionization. However excess N abundances are known
in high-$z$ QSOs (Hamann, 1999; Petitjean \& Srianand, 1999) and the
observed ratio could be due to the excess [N/C]. Other possibility for
getting the higher values of N(N~{\sc v}) could be the difference in
the covering factor for \nv and \civ (Petitjean \& Srianand, 1999).
ROSAT PSPC observations by Nandra et al.(1993) have confirmed the soft
X-ray absorption by ionized gas in NGC 5548.  Reynolds (1997) has
obtained $\tau_{O~{\sc vii}}$ = 0.24$\pm$0.04 and $\tau_{O~{\sc viii}}$
= 0.16$\pm$0.03 by fitting the observed ASCA spectrum using
$\Gamma~=~1.88\pm0.01$ and $N_H\le 2.2~10^{19}~cm^{-2}$.  The model
fits to the ASCA data investigated by George et al.(1998) suggest that
the warm absorbing material is completely covering the X-ray continuum
emitting region.  If we assume complete coverage then the estimated
column densities are $N({O~{\sc vii}})=9.3\times10^{17}$ and $N({O~{\sc
viii}})=1.2\times10^{18}~cm^{-2}$. Done et al. (1995) using the ROSAT
spectra obtained a lower limit on the electron density in the warm
absorbers to be 5$\times10^5~cm^{-3}$. Unlike UV spectrum the X-ray
spectra is of lower resolution and it is very difficult to get the
optical depth contributed by individual components, we detect in the UV
spectra.

\section {Results of the model calculations:}

\begin{figure}
\centerline{\vbox{
\psfig{figure=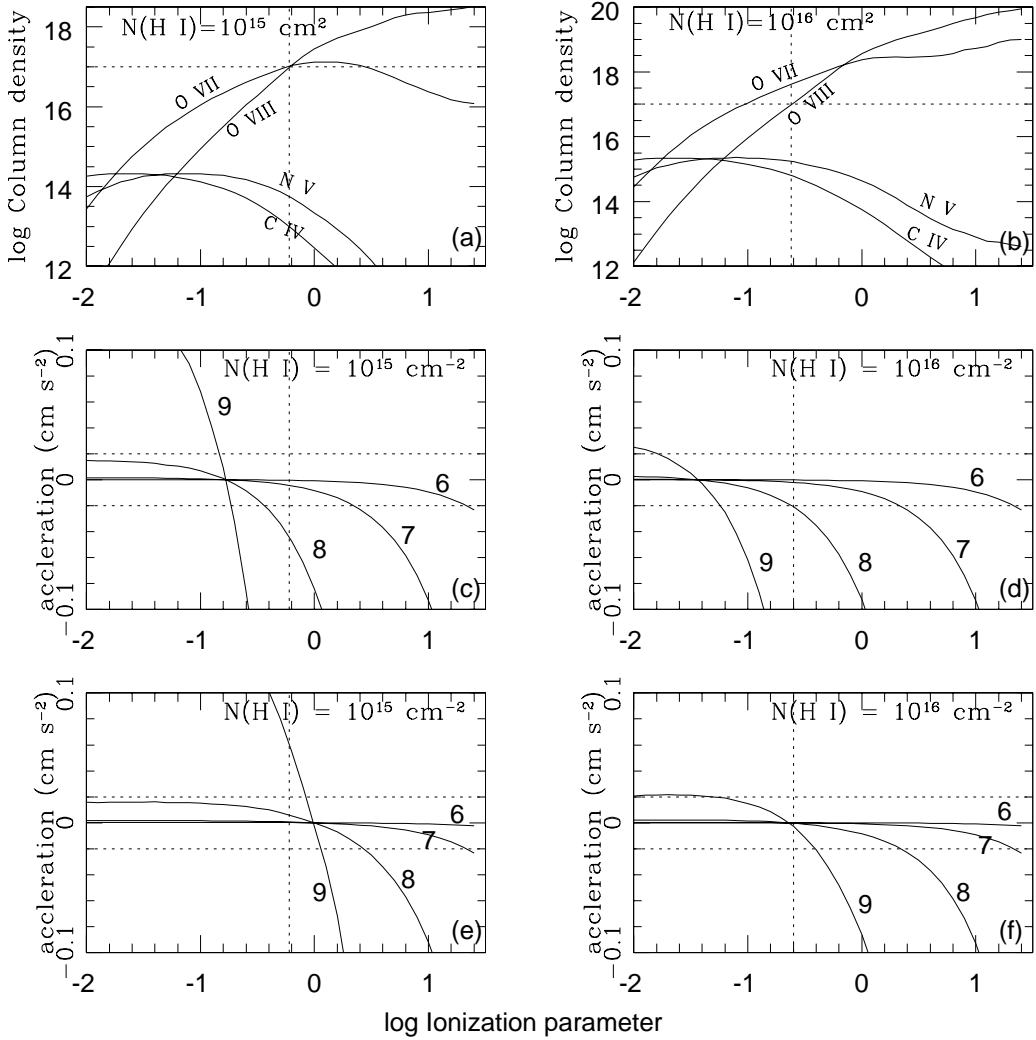,height=15.cm,width=15.cm,angle=0}
}}
\caption[]{ Results of estimation of radiative acceleration 
using photoionization models. Panels (a) \& (b) give the column
densities of various ions as a function of ionization parameter
for N(H~{\sc i}) = $10^{16}~{\rm and}~ 10^{15}~cm^{-2}$ respectively.
The net acceleration estimated for different models are given
in panels c($M=10^8M_\odot,~{\rm N(H~{\sc i})}=10^{16}~cm^{-2})$, 
d($M=10^8M_\odot,~{\rm N(H~{\sc i})}=10^{15}~cm^{-2}$), e($M=10^7M_\odot,~{\rm 
N(H~{\sc i})}=10^{16}~cm^{-2}$) and f($M=10^7M_\odot,~{\rm N(H~{\sc 
i})}=10^{15}~cm^{-2}$).  The horizontal dashed lines gives the observed
limit on the acceleration. The results are presented for four different
value of densities. The values of $log~n_H$ are marked close to the
corresponding curves. 
}
\label{fig3}
\end{figure}

We use the photo-ionization code cloudy (Ferland,1996) to get the
ionization conditions in a typical absorbing cloud.  We consider the
absorbing cloud to be plane parallel slab of uniform density with
chemical composition similar to that of solar composition. The ionizing
radiation is assumed to be similar to the one used by Dumont et
al(1998) and assume to shine on the cloud from one side. We run grids
of photo-ionization models covering -2.0$<log~U<$1.00 and stopped the
code when the column density of neutral hydrogen reaches a given
value.  In this work we discuss the results obtained when $N(H~{\sc
i})$ is 10$^{15}$ (model A) and 10$^{16}~cm^{-2}$ (model B)
respectively. The model results are plotted in Fig~{\ref{fig3}}. The
panels (a) and (b) give the column densities of various ionization
states that are detected in the associated system as a function of
ionization parameter. In the range of ionization parameter considered
here, the \cii column density is well within the upper limit we get
from the non-detection.  Also the resulting \nv column densities are
higher than that of \civ.  The horizontal lines in both the panels show
the lower limit on the column densities of O~{\sc viii} and O~{\sc vii}
so that the optical depth is equal to one tenth of the measured optical
depth.  Fig~\ref{fig3} suggests, in order for UV absorber to contribute
at least 10\% of the optical depth of Oxygen edges, that $log~U\ge-0.2$
(model A) and $log~U\ge-0.6$ (model B).  If only one  of the clouds
produce all the observed optical depth of ionized Oxygen edges then
$log~U\ge0.6$ (model A) and $log~U\ge-0.2$ (model B). In the case of
model A, such a cloud will not produce detectable amount of \civ and
\nv column densities as in the case of the sub-components in the
low-$z$ system (if the abundance ratios are solar).  However in the
case of model B, in a narrow range of ionization parameter, one can get
a consistent solution for the UV as well as "warm absorbers".

The column densities of various ionization states of H, He, C, N  and O,
resulting from our models, are stored in an array and are used to
estimate the equivalent widths of different transitions. We also store
the total column density of hydrogen resulting from the models. We
consider the Doppler broadening and natural broadening of lines (Voigt
profiles) while calculating the equivalent widths.  We estimate the
radiative acceleration due to line absorption using the column densities
of H~{\sc i}, He~{\sc i}, He~{\sc ii} and all possible ionization
states of O, N and C.  We use $H_o~=~75~kms^{-1}~Mpc^{-1}$,
$Q~=~3\times10^{54}~s^{-1}$, $q_o~=~0.5$, redshift of the AGN,
$z=0.0175$. The results presented in Fig~\ref{fig3} assume the
velocity dispersion to be 25~\kms. The acceleration due to gravity and
the radiation pressure are calculated assuming the mass of the central
engine to be $10^{8}M_\odot$  and $10^{7}M_\odot$. The net acceleration
as a function of ionization parameter are plotted for a range of
densities for the two cases (see panels c-f Fig.\ref{fig3}).  
The allowed range in the acceleration, from the observations, are given by the
horizontal dotted lines.  Positive values of the acceleration means the
radiative acceleration dominates and the negative values represent the
gravitational deceleration.

When mass of the central engine is $10^{8}M_\odot$, radiative
acceleration dominates when the ionization parameter $log~U\le-0.75$
(model A) and  $log~U\le-1.45$ (model B). Note that such a cloud will
have the optical depths of O~{\sc vii} and O~{\sc viii} less than 0.01
and could not make appreciable contributions to the observed optical
depths. Decreasing the mass of the central source will increase the
upper limit on $log~U$.  Note that when we consider $M=10^{7}M_\odot$ a
small range is allowed in $log~U$ where the cloud can contribute one
tenth of the observed optical depth of the Oxygen edges. However if all
the Oxygen edge optical depths are due to single cloud then the cloud's
motion will be dominated by the radiative acceleration due to line
absorption only when $M<10^7M_\odot$.  Such a low values of the mass
are not allowed by the observations.

When we consider $n_H=10^9~cm^{-3}$, observational constraints on the
change in velocity allows an extremely narrow range in $U$ when
$N(H~{\sc i}) \le 10^{15}~cm^{-2}$. Also this demands a near critical
balance between the radiative acceleration and the gravitational
deceleration, which requires extreme fine tuning. Also such a balance
can not be maintained when the ionizing continuum changes as in the
case of NGC 5548. Thus it is more likely that the density of the
absorbing cloud has to be less than $10^9~cm^{-3}$ when $N(H~{\sc i})
\le 10^{15}~cm^{-2}$.  This will mean the distance of the cloud from
the central source has to be higher than 20 lt days (for $log~U\le
-0.2$).  Transfer function derived from the HST monitoring data suggest
that the region of \civ emitting clouds streaches from 1 lt-day or
closer, out to 10 lt-day are further (Done \& Krolik, 1996). Thus the
locations of the clouds with N(H~{\sc i})$\le10^{15}~cm^{-2}$ will be the
outer regions of the BLR.  Note that the Voigt profile fits to the
components suggest that the $N(H~{\sc i})$ in individual components are
less than $10^{15}~cm^{-2}$. However, as discussed before, due to
unknown degree of saturation and covering factor the column density
derived will be a  lower limit only. The \lya line of the low-$z$
system is well resolved and, as can be seen from Fig~\ref{fig1}, the
lines do not show any signature of saturation (flat bottom) and it is
more likely that the $N(H~{\sc i})\le10^{15}~cm^{-2}$ in the individual
components. Thus the location of such a cloud will be between the BLR
and NLR. However the change in the velocity for these clouds are not measured due to poor S/N in the GHRS \& STIS data.

When we consider $N(H~{\sc i})=10^{16} ~cm^{-2}$, critical balance is
needed only for $n_H\ge10^{10}~cm^{-3}$. This means the distance of such a 
cloud from the central engine has to be higher than 2 lt days (for
$log~U\le 0.2$). Note that some of the components in the high-$z$
system could have such large column densities. However the \civ and \nv
profiles of these components suggest that these clouds cover more than
80\% of the BLR and have to be outside the BLR. Thus we believe
$N_H\le10^9~cm^{-3}$ even in these clouds.
 
In all our calculations we consider the effect of the continuum radiation
alone.
However if the absorbing clouds are outside the BLR then the line
photons from the emission line clouds could also contribute to the
radiative accleration. Specially \civ and \lya emission lines are very
strong and the excess contribution to the accleration due to these
lines could be appreciable. Also this effect will be larger for the
high-$z$ system. In order to keep our analysis simple we do not
include the effect of the emission lines in the radiative acceleration.

\section {Implications of line-locking in NGC 5548}

\begin{figure}
\centerline{\vbox{
\psfig{figure=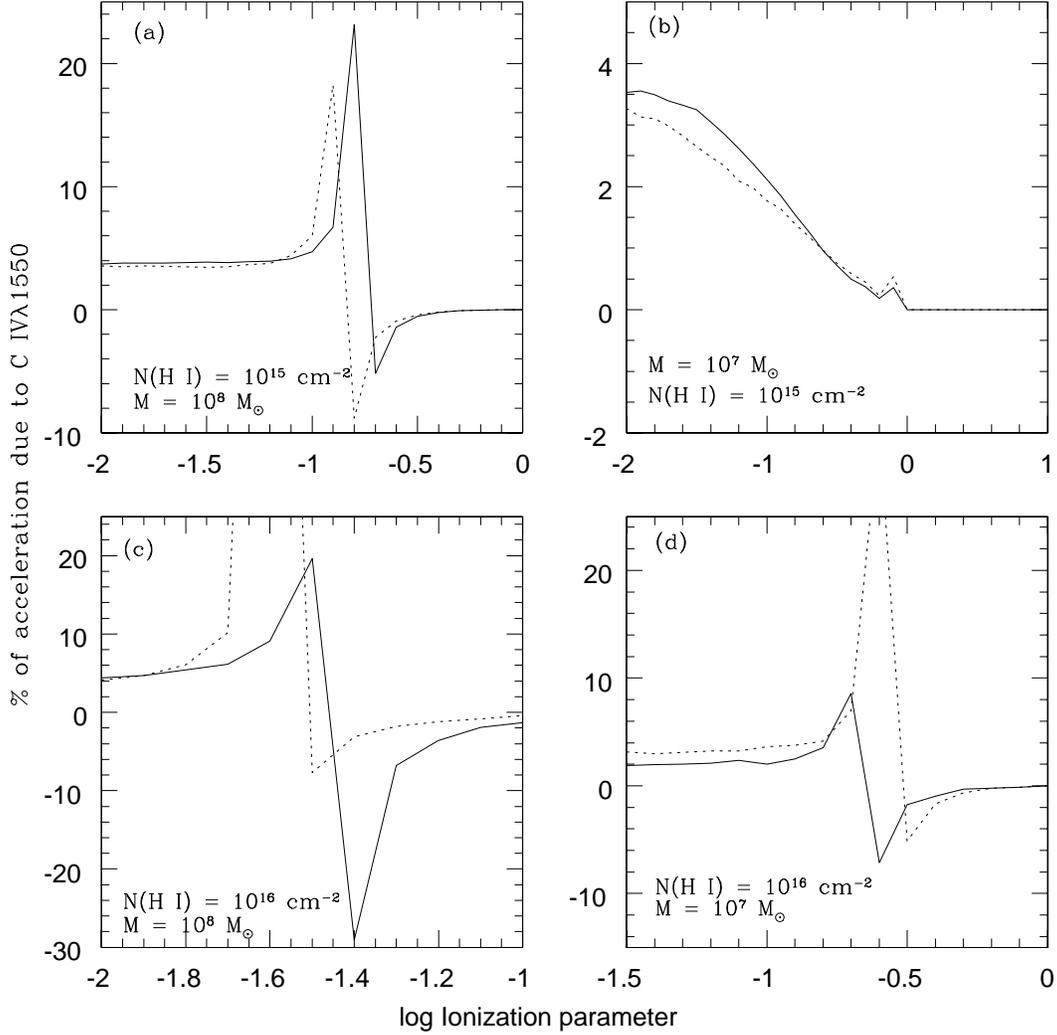,height=15.cm,width=15.cm,angle=0}
}}
\caption[]{Fractional acceleration contributed by the \civb
absorption as a function of ionization parameter. The solid
and dotted curves are the results obtained assuming the values of velocity
dispersion in the absorbing clouds to be 25 \kms and 10 \kms 
respectively. }
\label{fig4}
\end{figure}

In order for the line locking to occur the radiative acceleration
should contribute appreciably to the net acceleration of the cloud.  As
discussed before the locking among different sub-components in
absorption occures at the velocity difference corresponding to the
velocity splitting of \civ doublet.  This means the fractional
acceleration contributed by the \civb absorption line is important.
Since the locking occures at different velocities with respect to AGN
it is clear that the case we have to consider is non-steady state line
locking.  Also if the continuum source varies it will be easy to
preserve the locking under the non-steady state conditions. In
Fig.~\ref{fig4} we plot the fractional acceleration contributed by the
\civb absorption in different case under consideration.  As most of the
important transitions are saturated effect of velocity dispersion is
important.  In Fig~\ref{fig4}, we plot the results for two different
values of the velocity dispersion parameter in order to get the feel
for the dependence of radiative acceleration on our choice of velocity
dispersion.  Note that these curves are independent of the choice of
$n_H$. In the range of ionization parameter where the radiative acceleration
dominates \civb absorption contributes up to 4\% of the net acceleration.
The maximum contribution of the \civb line to the acceleration occurs
when there is a near balance between the gravitational acceleration and
the radiative acceleration. Locking at such ionization parameters are
very unstable against a small changes in the intensity of the ionizing
continuum. 
 
Let $a_1$ and $a_2$ are the net acceleration of two clouds along
our line of sight (with $a_2>a_1$).  Let $f_{\civ}$ be the fractional
acceleration contributed by the \civb absorption in cloud 2. At the
locking,
\begin{equation}
a_1~=~(1-f_{\civ})~a_2
\end{equation}

\noindent
If the ionizing continuum is same for both the clouds (i.e. cloud 1 is
optically thin to the continuum radiation) then the difference in
acceleration between two clouds can be achieved by changing $U$, $n_H$,
$N(H~{\sc i})$ and $N_H$. Note that the total acceleration is directly
proportional to the produce $Un_H$. In order to achieve the locking the
cloud which is getting locked (cloud 2) should be further away from the
source compared to the cloud which produces locking (cloud 1). This
demands,
\begin{equation}
U^an_H^a~>~U^bn_H^b,
\end{equation}
where superscripts $a$ and $b$ are used to represent the first and
second cloud respectively. This  condition suggests that if the
ionization parameter of the locking cloud is less than or equal to that
of the locked cloud (i.e $U^a\le U^b$) then the density of the locking
cloud should be higher than that of the locked cloud ($n_H^a>n_H^b$).
In order to understand the range of parameters one needs to get
the locking we consider the following cases.

\begin{itemize}

\item{\it case i}: If we assume  N(H~{\sc i}) for locking cloud is less
than or equal to that of the locked cloud, and if $n_H^a$ is less than
$n_H^b$ then $a_1$ will be always greater than $a_2$ in the range  of
ionization parameters relevant to this study. Thus in order for the
locking to occur $n_H^a<n_H^b$.

\item{\it case ii}:  If we assume  N(H~{\sc i}) for locking cloud is
greater than or equal to that of the locked cloud, then $a_2$ will be
much higher than $a_1$. However in order to produce locking at the \civ
velocity splitting, the difference in the acceleration between the
clouds should be equal to the acceleration of cloud 'b' due to \civb
absorption.  This condition suggest, in the range of $U$ relevant to
this study, $n_H^b$ will be equal to or slightly less than $n_H^a$
in order for the line-locking to occur at the \civ doublet splitting.

\end{itemize}

In the case of NGC 5548 the locked clouds (low-$z$) most probably have
a lower H~{\sc i} column density than the locking cloud (high-$z$). As
discussed above the $n_H$ of the low-$z$ system should be equal to or
less than that of the high-$z$ system in order to produce line-locking.
Though we do not have stringent limit on the change in the velocity of
the low-$z$ system, due to poor S/N, in the wavelength range of the
\civ absorption, the above condition suggests that the density of the
low-$z$ system is less than $10^9~cm^{-3}$. From Fig~\ref{fig3}  it is
clear that in the range of ionization parameter, where the radiative
acceleration dominates, the optical depth of O~{\sc vii} and O~{\sc
viii} in the clouds are much lower than the observed value. Though low
values of central mass can produce solutions with radiatively
accelerated clouds with appreciable optical depths of O~{\sc vii} and
O~{\sc viii}, the fractional acceleration contributed by \civb in such a
clouds are negligible. Thus the clouds which are showing line-locking
with \civ doublet splitting are most probably not responsible for the
observed ionized Oxygen edges. 
 
\section {Discussion:}

HST observations of nearby AGNs suggest that $\sim50\%$ of the seyfert
1 galaxies do show associated absorption systems (Crenshaw et  al., 1998).
Invariably, in all cases \zabs~ is less than \zem, suggesting a component
of out-flowing, highly ionized, optically thin gas, close to the centers
of seyfert galaxies. The strong absorption components have cores that
are much deeper than the continuum flux levels, indicating that the
regions responsible for these components lie completely outside the
BLR. This result is supported by the derived covering factors for few
systems (NGC 3516, NGC 3783, NGC 4151, NGC 5548, Mrk 509) using the
high resolution data (Crenshaw et al., 1998), and the constancy of the
wavelength of the absorption lines over a period of few years seen in
few systems (NGC 5548 (this work) and NGC 4151 (Weymann et al. 1997)).
Presence of signatures of line-locking, in NGC 5548, suggest
that the radiative acceleration plays (or played) a vital role
in accelerating the clouds. Note that even
in the case of NGC 4151 some of the narrow components seem to have 
velocity splitting close to the \civ doublet splitting (see Fig 5.
of Weymann et al. 1997).

The 1993 monitoring campaign on NGC 5548 (Korista et al. 1996) revealed
a red/blue symmetry between the blue and red line core light curves of
the \civ emission. However no such relationship is seen between the
blue and red wing light curve suggesting a possible radial motions for
emitting clouds with large velocities. Done \& Krolik (1996), noted
that the response function for the red and blue cores are nearly
identical, and the red wing response has a strong peak at small lags
but the blue wings response raises only very slightly toward zero time
delay. This indicates, if we assume all the emitting cloud to have a
positive response, that there is more material near the line of sight
and/or at small radii that is traveling away from us than is traveling
towards us.

Detail studies of UV emission lines in NGC 5548 provide some indirect
evidence for the  presence of optically thin line emitting clouds.
Sparke(1993) finds that gas with negative response is required, at
small time delays, to explain the shape of the cross-correlation
function for several lines in the case of NGC 5548. Koen(1993) finds
evidence for negative instantaneous response of \siiv and C~{\sc ii}
lines to the UV continuum variations.Various implications of the
presence of such optically thin components are discussed by Shields et
al. (1995). Observations discussed here clearly suggest that there is a
component of optically thin gas, which are most probably present just
outside the BLR, with net outward velocity. As the densities are high the
cloud will respond to the continuum variations instantaneously. The
\civ line emission from these clouds will have negative response to the
continuum variations. If the volume filling factor of such clouds are
high then they are expected to contribute appreciable to the total \civ
emission.  The slow response towards the zero time lag in the case of
blue wing could be due to the instantaneous negative response of small
fraction of optically thin clouds close to our line of sight. If these
clouds are far away from the central continuum source and have radial
outflows, like the individual components in the absorption systems,
then they will affect only the blue wing response function at short
time delays. However these clouds can contribute appreciable amount to
the response function of the red wing at higher time delay.
 
It is interesting to note that the response functions derived for
total \civ as well as red wing of the \civ emission show minima
(even negative response) at a delay time of $\sim 20$ days (Done
\& Krolik, 1995). Similar result is also obtained by Wanders et al. (1995)
using different numerical method.  They interpreted this as an
evidence for bi-conical structure in the emission line region. However
this result is also consistent with our expectation in the case of 
presence of radially out-flowing optically thin gas in the outer regions of the BLR. Though there are models of BLR with optically thin as well as optically thick  clouds are discussed in the literature (Goad et al. 1995),
optically thin clouds are considered very close to the central source. It
will be an interesting and important exercise to build models of
BLR with optically thin clouds in the outer regions as suggested by
the absorption line studies. 

\section {Summary}

We study the multiple velocity component structure in the associated
absorption system in NGC 5548 using HST high resolution spectra. 
%We show that most of the narrow components in the associated systems are
%line-locked with velocity separation equal to that of the \civ doublet
%splitting. 
Most of the narrow components in the associated systems show the
signature of line-locking with velocity separation equal to \civ
doublet splitting.
Using the spectra observed at two different epochs separated
by $\sim 2$ yrs we estimate the upper limit on the acceleration of
individual clouds.  We construct grids of photo-ionization models and
estimate radiative acceleration due to line absorption by various
transitions. Using the constraints on the ionization parameter,
acceleration, covering factor and the conditions to be satisfied for
the line-locking, we show that (a) the density of the cloud is less
than 10$^{9}~cm^{-3}$, (b) the clouds are situated out side the BLR,
(3) the clouds involved in the line-locking can not be responsible for
the observed ionized Oxygen edges in the soft X-ray. We discuss the
implications of the presence of optically thin clouds in the outer
regions of the BLR. Lack of information on the ionization conditions of
the clouds prevents us from making a more realistic models of the
line-locked clouds.  We believe, future UV observations of NGC 5548
using FUSE will provide better constraints on the ionization parameters
of the individual components, which can be used to get a more realistic
models.

%%%%%
\newpage
\newpage
\newpage
\newpage

\end{document}